# Towards
# A MODEL FOR COMPUTING
in European Astroparticle Physics

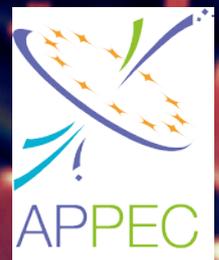

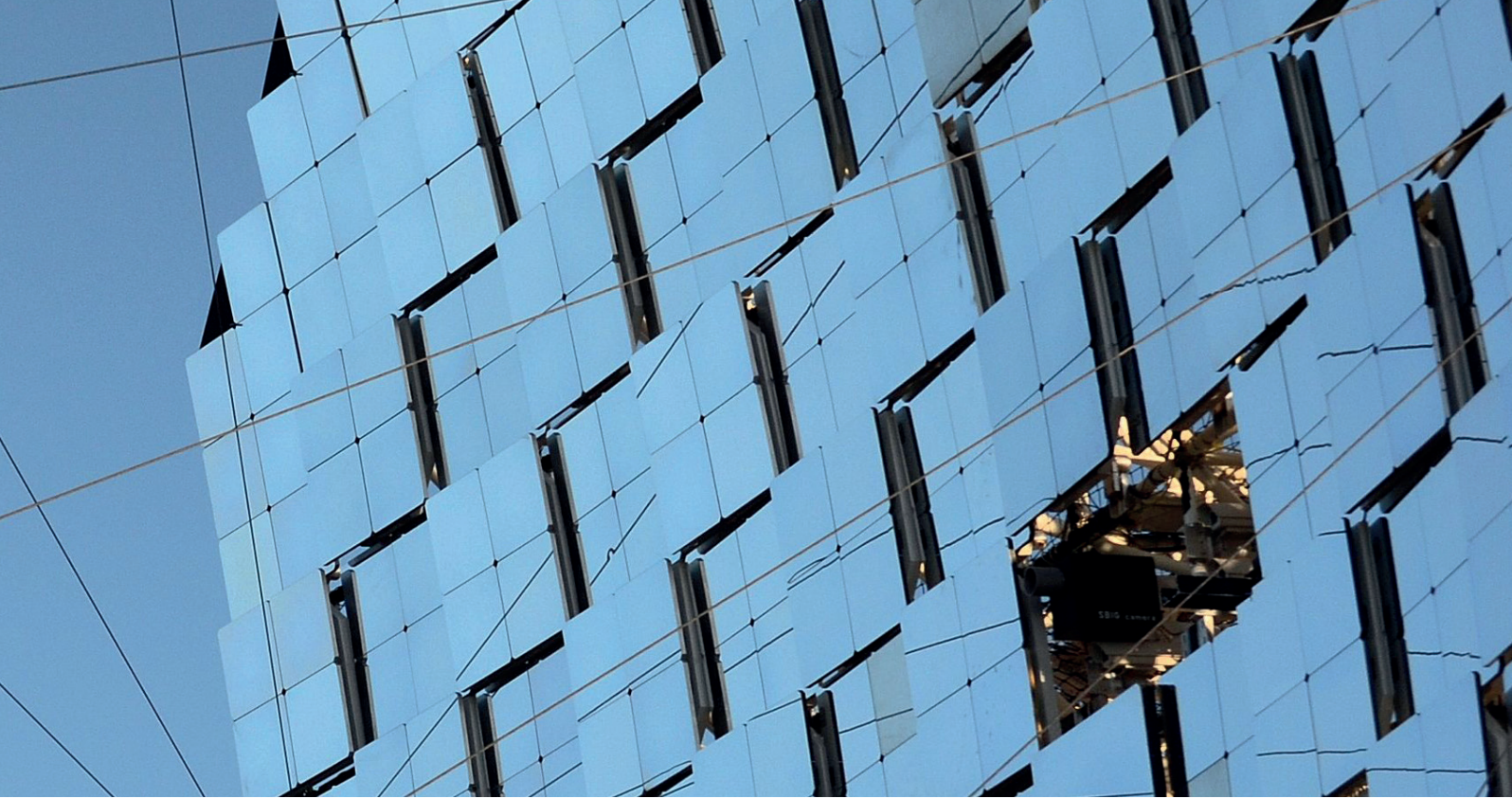

# Contents

# Executive summary

Current and future astroparticle physics experiments are operated or are being built to observe highly energetic particles, high energy electromagnetic radiation and gravitational waves originating from all kinds of cosmic sources. The data volumes taken by the experiments are large and expected to grow significantly during the coming years. This is a result of advanced research possibilities and improved detector technology.

To cope with the substantially increasing data volumes of astroparticle physics projects it is important to understand the future needs for computing resources in this field. Providing these resources constitutes a larger fraction of the overall running costs of future infrastructures.

The document presents the results of a survey made by APPEC with the help of computing experts of major projects and future initiatives in astroparticle physics, representatives of current Tier-1 and Tier-2 LHC computing centers, as well as specifically astroparticle physics computing centers, e.g. the Albert Einstein Institute for gravitational waves analysis in Hanover. In summary, the overall CPU usage and short-term disk and long-term (tape) storage space currently available for astroparticle physics projects' computing services is of the order of one third of the central computing available for LHC data at the Tier-0 center at CERN. Till the end of the decade the requirements for computing resources are estimated to increase by a factor of 10.

Furthermore, this document shall describe the diversity of astroparticle physics data handling and serve as a basis to estimate a distribution of computing and storage tasks among the major computing centers. Given that astroparticle physics projects depend on certain site conditions and are mostly placed at remote locations often without direct access to high-bandwidth network each project is required to develop its own solution to transfer project data to computing centers for a further data processing. Essentially all astroparticle physics data are processed in more than one computing center, either all in Europe or in combination with centers in other regions, which is an important fact concerning data safety and the organization of a shared access to the data.

Astroparticle physics data can be categorized as events, time-series and images. Each type of data requires a different type of analysis method. The organization of services and resources required for data analysis is different from community to community. Generally speaking, projects generating event type data are currently making use of existing resources provided by HTC centers federated through the computing-grid middleware, the collaborations of gravitational wave experiments have developed dedicated solutions and built up resources to process recorded time-series, and data reduction of dark energy surveys follows standard procedures developed in astrophysics.

The diversity of data and its analysis makes astroparticle physics a formidable test-bed for new and innovative computing techniques concerning hardware, middleware, analysis software and database schemes. Tier-2 scale computing centers can play an important role as hubs of software development, visualization, general interfacing, coordinated design, and public outreach.

In addition to the discussion on how to cope with the future computing requirements of the individual projects it is important to discuss with the astroparticle physics community how to make data formats of results compatible. This is especially important to ease multi-messenger analyses and beneficial for organizing the access to the data, possibly by one common interface. Access policies developed in astrophysics for data taken at large astronomical observatories or by satellites can also be adopted for astroparticle physics data; after a defined proprietary period – suitable for a scientific analysis by the Principle Investigator – data are made publicly available via archives. It should be considered to organize such an open access to the data in the frame of the Virtual Observatory, a common interface to astrophysical data developed by the astronomy community.

# Table of projects

# Table of abbreviations

**Cloud**  Cloud computing services:
- SaaS (Software as a Service) – software is available on remote computers, and data can be either local or remote
- PaaS (Platform as a Service) – complete software platforms are available on remote Datacenters
- DaaS (Data as a Service) – data are available on remote computers, and can be accessed either locally or remotely
- HaaS (Hardware as a Service) – user has both data and software, and send these to remote computers to run the jobs
- IaaS (Infrastructure as a Service) – similar to Grid Computing but resources are used on demand without the need to pledge them

**CORSIKA**  COsmic Ray SImulations for Kascade
http://www.ikp.kit.edu/corsika

**DSP**  Digital Signal Processor

**EAS**  Extensive Air Shower: cascade of ionized particles and electromagnetic radiation produced in the atmosphere when a primary cosmic ray enters the atmosphere

**EGI**  European Grid Infrastructure
http://www.egi.eu/

**ESA**  European Space Agency

European Southern Observatory

**EMBL**  European Molecular Biology Laboratory
http://www.embl.org/

**Flash-ADC**  Flash analog-to-digital converter

**FPGA**  Field Programmable Gate Array

**HEAVENS**  High-Energy Astrophysics Virtually ENlightened Sky: a High Energy Astrophysics specific interface/archive and add-on to the Virtual Observatory (VO)
http://isdc.unige.ch/heavens/

**HepSpec 2006**  high-energy physics wide benchmark for CPU performance
http://w3.hepix.org/benchmarks

**HPC**  High-Performance Computing

**HTC**  High-Throughput Computing

**ISDC**  INTEGRAL Science Data Center
http://www.isdc.unige.ch/

**ISS**  International Space Station

**kHS06**  processing resources in thousands of HepSpec 2006

**LHC**  Large Hadron Collider

**LSC Data Grid**  LIGO data Grid is the combination of LIGO Scientific Collaboration computational and data storage resources with grid computing middleware
https://www.lsc-group.phys.uwm.edu/lscdatagrid/

**NAF**  German National Analysis Facility: serving the nat. particle physics community
https://naf-wiki.desy.de/Main_Page

**SPEC2006**  industry-standardized, CPU-intensive benchmark suit http://www.spec.org

**VO**  Virtual Observatory: collection of inter-operating data archives and software tools utilizing a web-based environment for astronomical research
http://www.virtualobservatory.org/
http://www.euro-vo.org

**WLCG**  Worldwide LHC Computing Grid
http://wlcg.web.cern.ch/

# 1. Introduction

The last decade witnessed the intensive construction of large astroparticle physics experiments and detectors. Towards the end of the decade the projects passed from the noise hunting regime to the generation of large sets of data, whose production needs large computing resources, intensive simulation and large storage space. Furthermore, the data need to be cast in formats and be accompanied by software permitting wide accessibility as well as correlation with information generated by other observatories of astrophysical or cosmological type.

The domain will experience in the coming decade exponential growth of data produced; participating in the overall forthcoming "data tsunami" and presenting major challenges in terms of storage, computation and long term preservation.
This document results from discussions of the computing models for astroparticle physics projects. Three workshops have been organized on the following aspects of this model:
1. Grid and Virtual Observatory (October 2010, Lyon)[1]
2. Middleware (May 2011, Barcelona)[2]
3. Hardware (May 2012, Hanover)[3]

Furthermore, the replies of selected projects to a questionnaire have been analysed and discussed in a summing up meeting of the working group on July 16, 2013 in Paris. In a meeting in Geneva (October 29, 2013), an intermediate draft of this document was presented and discussed. A final meeting to scrutinize the numbers for computing needs and define the final document structure took place in Bologna (April 14/15, 2014). The needs and requirements have been refined in interaction with the scientific community. The document is structured as follows:
- Section 2 describes in detail the current status and the future needs for computing and data storage by astroparticle physics projects.
- Section 3 summarizes the astroparticle physics related data services and computing required from computing centers.
- The access and the scientific use of the data as seen from the perspective of the user/researcher are then provided in Section 4.
- A concluding summary is given in Section 5.
- A summary of quantitative numbers of computing and storage needs broken down by projects is provided in the Annex.

[1] https://indico.in2p3.fr/event/3845/
[2] http://indico.cern.ch/event/134280/
[3] http://indico.cern.ch/event/159120/

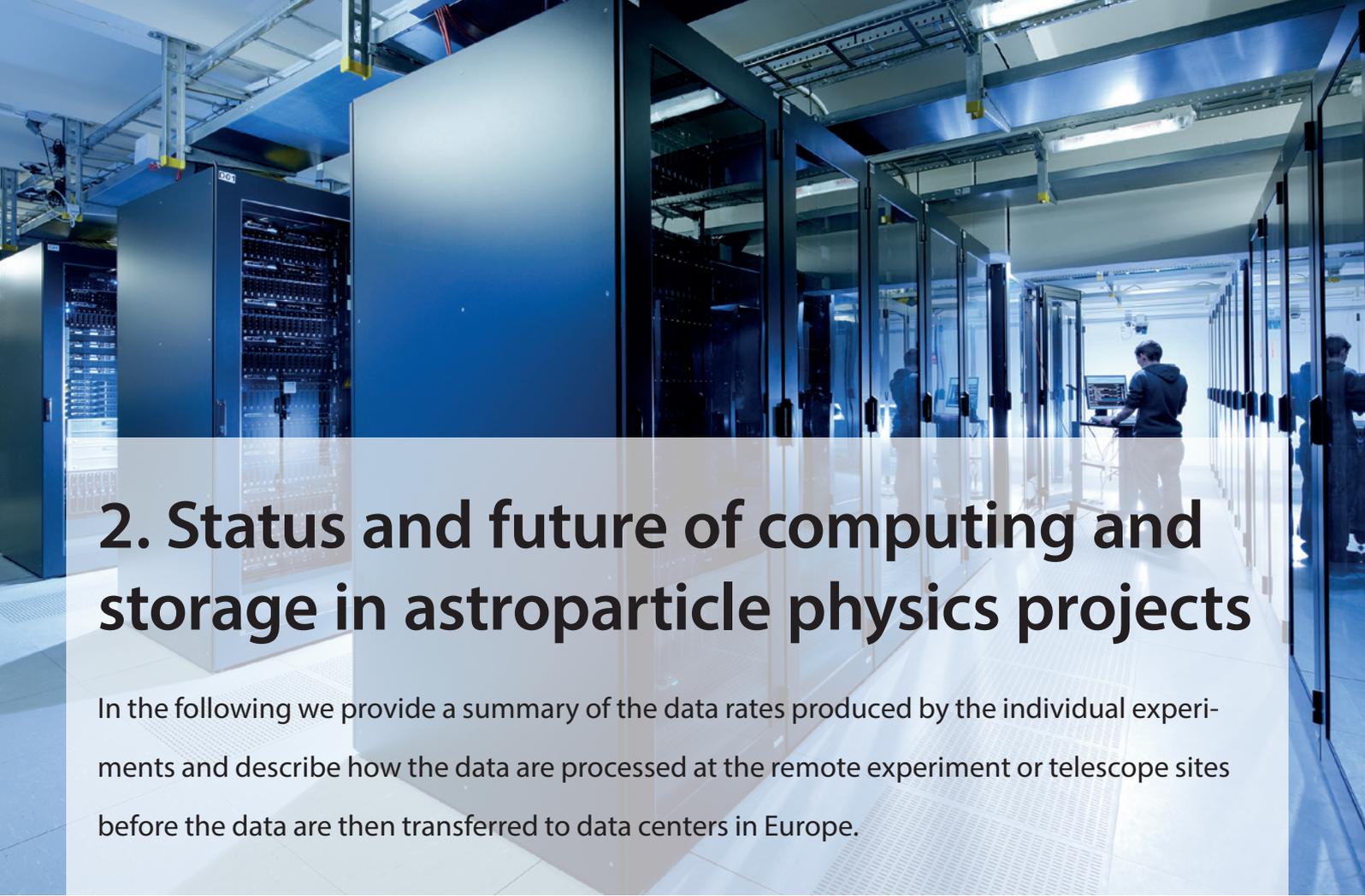

# 2. Status and future of computing and storage in astroparticle physics projects

In the following we provide a summary of the data rates produced by the individual experiments and describe how the data are processed at the remote experiment or telescope sites before the data are then transferred to data centers in Europe.

## 2.1 Data rates

Data rates delivered by currently running and planned research infrastructures are quite diverse, ranging from 10–20 GBytes/day from satellite experiments, to 100 GBytes/day (after reduction) for ISS based experiments or ground-based cosmic ray experiments, to a few TBytes/day for e.g. gravitational wave experiments, to 100 TBytes/day (before on-site reduction) for future high-energy gamma-ray observatory (CTA) and of the order of 15 TBytes per day for large ground-based telescopes (LSST). Altogether, the field of astroparticle physics therefore produces a data rate of a few PBytes/year already today, which is comparable to the current LHC output. To transfer this amount of data from the remote sites to the data centers at least Gb/s links are required. For space experiments the telecommunication system is the limiting factor of how much scientific data can be obtained. High bandwidth satellite links are needed. For Euclid the data transfer will be performed at a maximum data rate of 850 Gb/day (compressed). This number translates in a factor of 2–3 higher scientific data throughput.

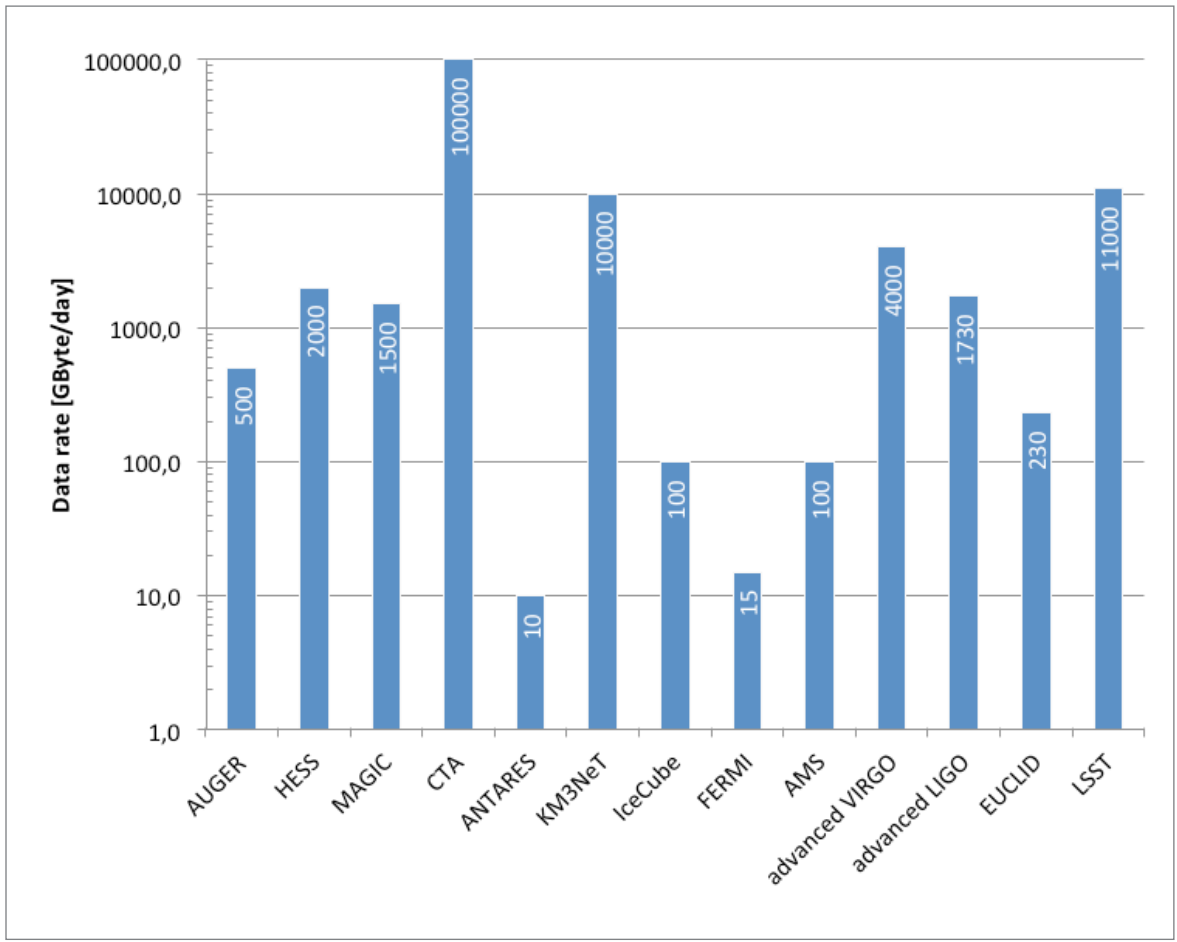

**Figure 2:** Data rates produced by the individual experiments.

## 2.2 Data acquisition and first data processing on site

Generally speaking, all experiments are recording housekeeping data describing the observation run, status of the detectors, cameras, or telescopes and data required for calibration together with the actual scientific data. Common for all experiments is a high precision timing system. To constantly verify that an experiment or telescope is fully functional and to check the quality of observations, real-time calibrations and a pre-processing of the data and/or other forms of data characterization are performed. In the following we present three examples of typical astroparticle physics experiments.

In modern Cherenkov telescopes, which are typical examples of experiments taking data of very fast events, the analog signal from the individual camera photosensors is digitized by a Flash analog-to-digital converter (Flash-ADC) and passed to a Field Programmable Gate Array (FPGA) based trigger logic deciding which events shall be recorded by the data acquisition system. Similar systems are implemented in all experiments dealing with very fast events such as low- and high-energy neutrino telescopes as well as cosmic ray experiments. In addition large on-site ICT infrastructures based on standard computer farms and networks will be used for data acquisition and data pre-processing.

Gravitational wave interferometers are experiments producing time-series data. Essential for these experiments is a robust and reliable data acquisition system capable of acquiring signals over periods of several months with as little interruption as possible. The typical data acquisition system of a gravitational wave interferometer consists of digital signal processor (DSP) based front-end readouts that are synchronized with the timing system. The raw data produced by the various front-end readouts are then collected into a structured format. This format has been commonly agreed on for all gravitational wave interferometers.

Dark energy surveys are carried out at large astronomical telescopes to map the entire visible sky deeply in multiple colors over a longer period of time. The astronomical images are stored together with calibration frames in a common data format. After data transfer to a computing center the scientific information will be reduced from the data to perform a time-lapse digital imaging of faint astronomical objects across the entire sky.

## 2.3 Data transfer to Europe

It is interesting to note that some astroparticle physics experiments are installed in very remote places where a direct link to the Internet is not available or not capable to take the load of transferring the entire data. Data are then copied on tapes or disks and transferred by mail or courier. Figure 3 shall visualize the complexity of the data flow from the individual experiments to the European data centers.

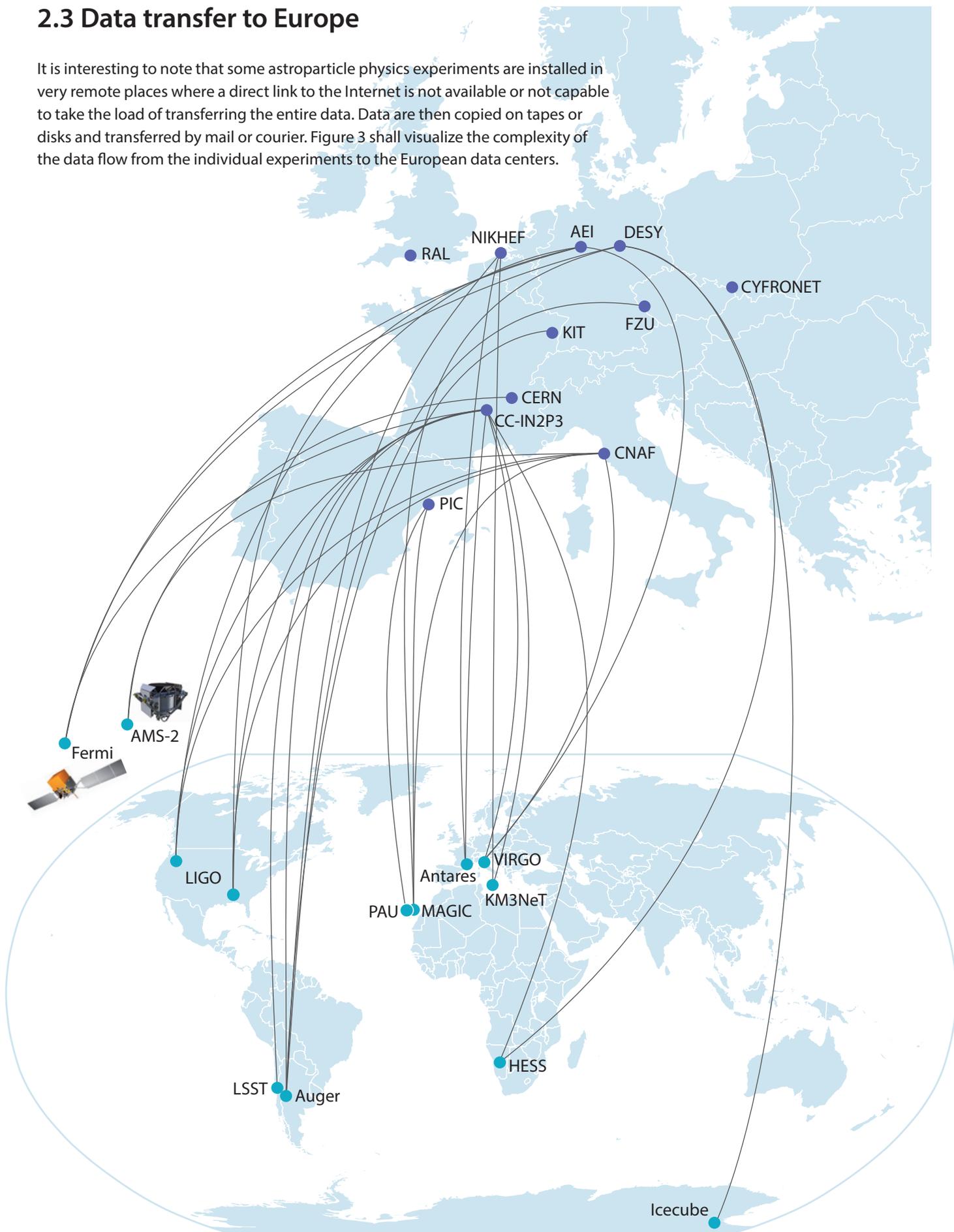

**Figure 3: Sketch of the current data flow from experiment sites to European data centers**

# 2.4 Resources for computing and storage in astroparticle physics projects

To assess the future computing requirements of astroparticle physics projects a questionnaire has been developed and distributed to the collaborations' computing experts. The intension of the questionnaire was to collect numbers of existing and required computing power and storage needs, to ask for special expertise or need in software and hardware, and to obtain details on data acquisition, processing, and data analysis. The data produced by the distinct experiments are different but can be characterized as follows:
- events: time stamped energy and direction data recorded e.g. by observatories detecting charged cosmic rays or high and low energy photons and neutrinos,
- time-series: time stamped signal recorded e.g. by gravitational wave antennas,
- images: two-dimensional data taken by astronomical telescopes e.g. to carry out dark energy and dark matter studies.

In the following, the needs in CPU power as well as short and long term storage of major current and future astroparticle physics projects, classified along the 3 categories above, are presented.

In the questionnaire the CPU performance was requested to be given in kHS06, a unit that results from a combination of the SPEC2006 suite floating point and integer benchmarks. This unit is commonly used in particle physics computing to specify requirements and resources; 1 kHS06 equals to the power of 100 CPU cores as of 2012. In experiments for gravitational wave detection the computing power is however measured in units of Tflop. To make the data comparable, 1Tflop (double-precision peak) has been converted into 1 kHS06. Furthermore, for a visualization of the total demand in computing power and storage needs the values have been related to the capability of the LHC Tier-0 center at CERN[4], which in 2012 had 65,000 cores installed and provided disk storage of 30 PByte and archive storage space of 30 PBytes. An "LHC Tier-$0_{2012}$ unit" thus denotes the capability of 65,000 CPU cores in 2012 and short and long term storage of 30 PBytes.

The results for the individual experiments are provided in three tables in Annex I. In Figure 1 the current use and future requirement for astroparticle physics experiments is provided in units of LHC Tier-$0_{2012}$. As can be seen from Figure 1, the current overall CPU usage sums up to 0,5 LHC Tier-$0_{2012}$ units. From 2015 on the CPU usage is expected to grow up to a value of about 3 until 2020. In the same time the short term disk storage shall grow from currently 0,25 to 1,5 LHC Tier-$0_{2012}$ units by 2020. The corresponding growth on long term (tape) storage will be from 0,25 to roughly 5 units. Figure 1 clearly demonstrates that with the start of the next generation of experiments around 2015 there is an increasingly demand in computing power and storage space. It is worthwhile to mention that the update report of the computing models of the WLCG and the LHC experiments[5] April 2014 quotes an increase from 2012 to 2017 by a factor of 3 and 2, respectively for Tier-0 computing and disk storage; the WLCG is the LHC Computing Grid of worldwide distributed computing infrastructure arranged in Tier data centers.

[4] http://press.web.cern.ch/press-releases/2012/05/cern-awards-major-contract-computer-infrastructure-hosting-wigner-research
[5] CERN-LHCC-2014-014/LCG-TDR-002: http://cds.cern.ch/record/1695401/

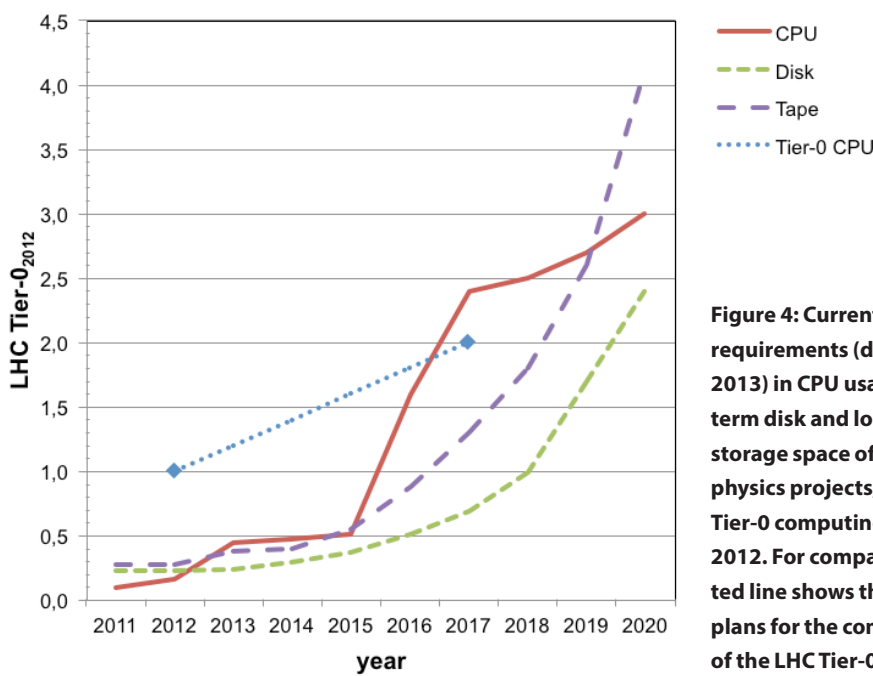

Figure 4: Current use and future requirements (data collected in 2013) in CPU usage and short-term disk and long-term (tape) storage space of astroparticle physics projects, in units of LHC Tier-0 computing and storage in 2012. For comparison the dotted line shows the extension plans for the computing power of the LHC Tier-0.

# 3. Data and computing services for astroparticle physics

Table 1 provides an overview of the European data centers managing astroparticle physics data. The last column entitled "Particle Physics Data" indicates whether the Computing center does also function as Tier center in the WLC.

| Computing Center | Country | Astroparticle Data |  |  |  |  |  |  |  |  |  |  |  |  |  |  | Particle Physics Data |
|---|---|---|---|---|---|---|---|---|---|---|---|---|---|---|---|---|---|
|  |  | AUGER | HESS | MAGIC | CTA | ANTARES | KM3NeT | IceCube | FERMI | AMS | VIRGO | LIGO | PAU | SNLS | EUCLID | LSST | Other | LHC |
| FZU | Czech Republic |  | X |  |  |  |  |  |  |  |  |  |  |  |  |  |  | Tier 2 |
| CCIN2P3 | France | X | X |  | X | X | X |  | X | X | X | X |  | X | X | X |  | Tier 1 |
| AEI | Germany |  |  |  |  |  |  |  |  | X |  | X | X |  |  |  |  |  |
| DESY | Germany |  | X | X | X |  |  | X | X |  |  |  |  |  |  |  | X | Tier 2 |
| KIT | Germany | X |  |  |  |  |  |  |  |  |  |  |  |  |  |  | X | Tier 1 |
| CERN | Transnational |  |  |  |  |  |  |  |  | X |  |  |  |  |  |  |  | Tier 0 |
| INFN-CNAF | Italy |  |  | X | X |  | X |  | X | X | X | X |  |  |  |  | X | Tier 1 |
| NIKHEF | Netherlands | X |  |  |  | X | X |  |  |  | X[1] |  |  |  |  |  | X | Tier 1 |
| NDGF | Nordic |  |  |  |  |  |  |  |  |  |  |  |  |  |  |  |  | Tier 1 |
| CYFRO-NET | Poland |  |  |  | X |  |  |  |  |  |  |  |  |  |  |  | X | Tier 2 |
| PIC | Spain |  |  | X | X |  |  |  |  |  |  |  | X |  |  |  | X | Tier 1 |
| RAL | UK |  |  |  |  |  |  |  |  |  |  |  |  |  |  |  | X | Tier 1 |

[1] data analysis support only

Table 1: Summary of European data centers managing astroparticle physics data

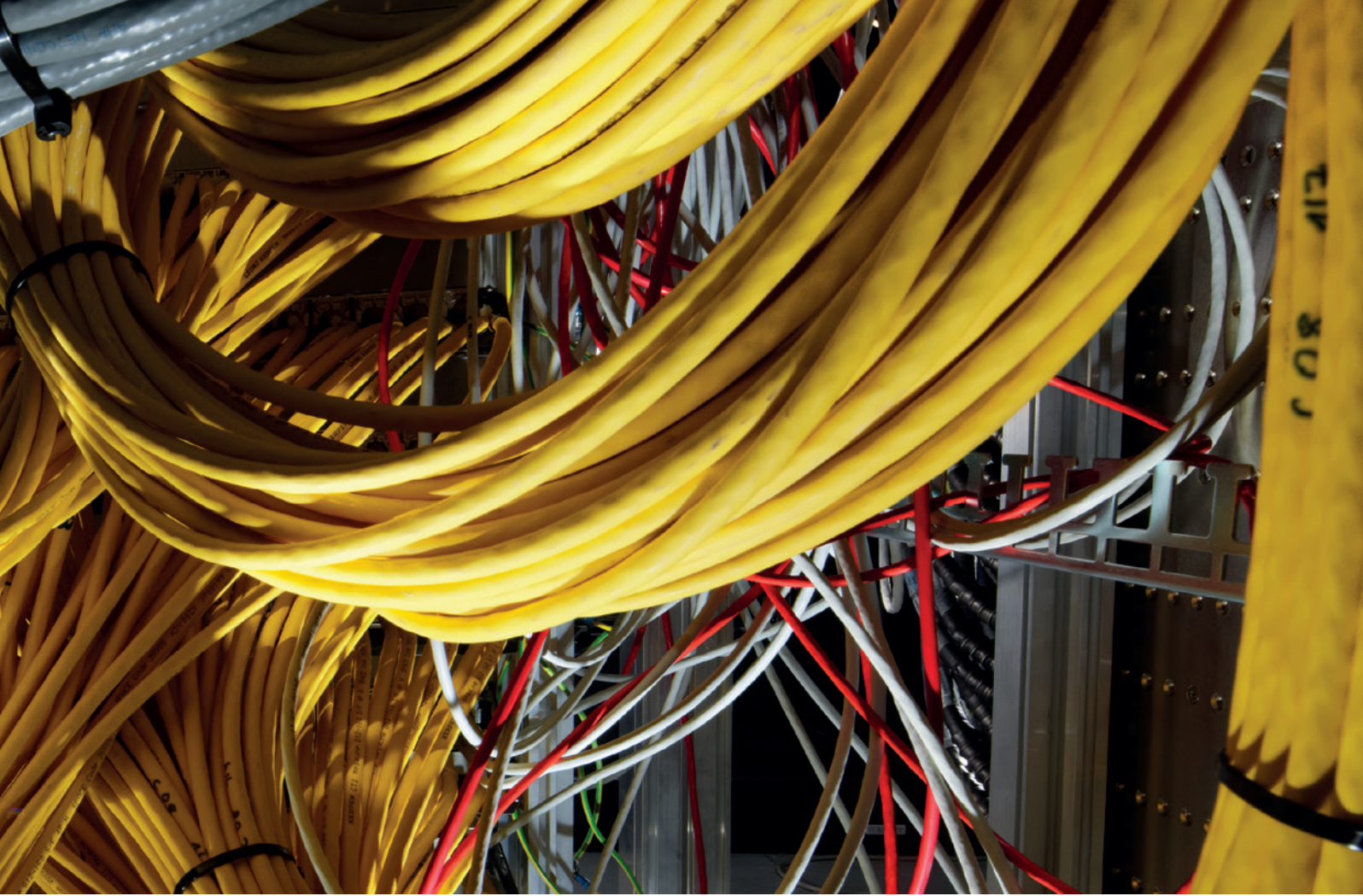

The domain of astroparticle physics faces a true challenge concerning the data to be treated. Not only, as in most areas of research today, we are seeing an immense growth in data volume, but also an increase in complexity and diversity. Thus, we have to deal with data sets that are too large and complex to manipulate or interrogate with standard methods or tools. In addition to this challenge, astroparticle physics has much to gain from the multi-messenger approach. In order to derive the highest scientific return from the data, it is necessary to cross-correlate data not only from experiments of the same type (e.g. from several neutrino detectors) but also across different domains. For example, a gamma-ray burst detected in hard X-rays is used as a trigger to search for a signal in data from neutrino and gravitational-wave experiments, and simultaneously a multi-wavelength analysis is performed including data from a large variety of observations ranging from the radio to the very high energy domain. These tasks must use data of different format, of different time, energy, and spatial resolution, and of different size. The real challenge is to bring together communities with different background, finding a common technical language in order to collaborate. Further complexity arises when communities apply distinctive rules concerning data rights and access.

Many of the astroparticle physics experiments rely heavily on simulations. The response of an experiment to its environment has to be simulated. Here we use more than ever computer simulations in addition to laboratory experiments, where the latter become impossible with ever-growing size and complexity of the observatories and detectors, such as the Cherenkov Telescope Array (CTA).
Thus, improved techniques have to be developed in order to handle the substantially increasing number of data produced by future astroparticle physics experiments and to allow for more advanced analyses with user-friendly data interfaces. Data preservation is another important task related to managing scientific data including those from astroparticle physics projects. Securely storing these data in a suitable number of repositories and ensuring their long-term readability is essential. Note that an intelligent geographical distribution of data and repositories may also help to minimize data traffic. In view of long-term data preservation a complete documentation of the data and its analysis is necessary. Furthermore, it is important to allow reanalyzing data sets of past experiments independent of the future availability of current computer hardware. The current approach to preserve complete analysis frameworks is the virtualization. On future computing hardware virtual machines replicating the operation system and analysis software used to analyse data of a certain past experiment might become a standard solution.

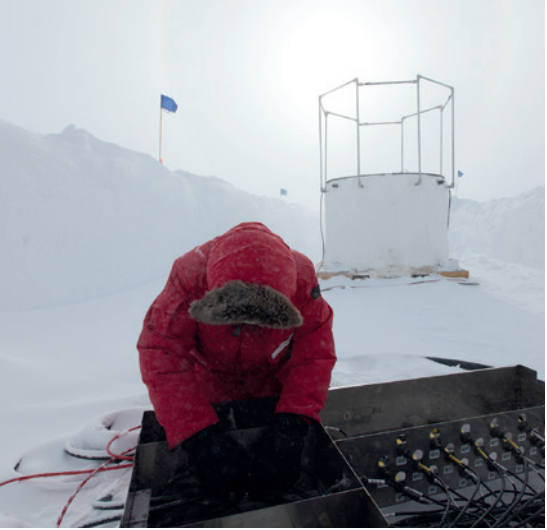
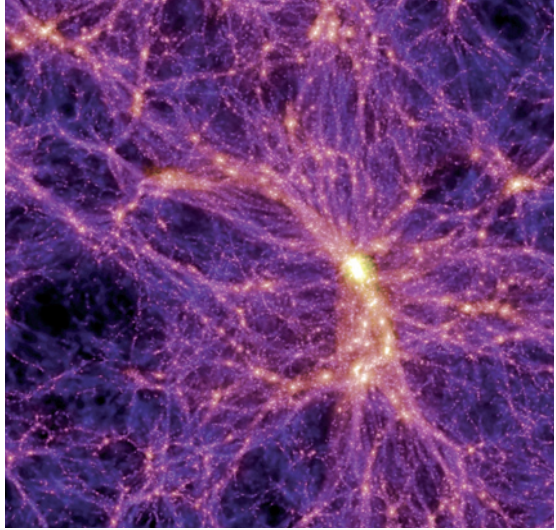
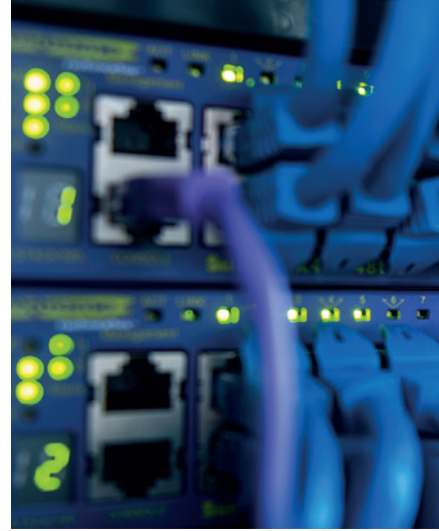

## 3.1 Data analysis, simulation and modelling

Computing services required by astroparticle physics projects mainly concern the analysis, simulation and modelling of data. The analyses to be performed for recorded astroparticle physics data of the respective types of data (events, time-series, and images) and projects are substantially different. They include calibration tasks, statistical methods to extract a scientific signal and often go hand in hand with simulations. Information on analysis frameworks currently used by project collaborations is provided in Section 4.

Simulations required for the data analysis are often more CPU-consuming than the capture and analysis of real data, in particular for the high energy cosmic ray experiments (charged particles, photons, neutrinos) and the gravitational wave antennas.

Experiments dealing with extensive air showers (EAS) induced by energetic cosmic rays, gamma rays or neutrinos require detailed simulation and reconstruction of the evolution and properties of the EAS in the atmosphere of the Earth. The standard tool for such simulations is CORSIKA (COsmic Ray SImulations for Kascade). CORSIKA was originally designed in 1989 for the KASCADE experiment at the Karlsruhe Institute of Technology. It is a numerical code using Monte Carlo technique to simulate the interactions and decays of particles and photons involved in the development of EAS up to energies of some 100 EeV. Optional code enables the production of neutrinos, Cherenkov radiation, or radio emission.

The computing requirements for performing EAS simulation and reconstruction are quite demanding and require large resources, both CPU time and storage. Depending on the type of primary particle and its energy such EAS simulations can last between CPU hours and CPU years; towards higher energies an increasing number of secondary particles and their interactions have to be computed. Only recently the use of GPUs (Graphical Processing Units) has permitted to consider a full and detailed simulation of the highest energy showers. To obtain statistically significant results at least a few thousands of events for each energy and particle type must be simulated. CORSIKA output files typically have sizes in the range of several GBytes up to several hundred GBytes.

Gravitational wave experiments' data analysis is the other astroparticle physics domain with a substantial demand in computing services. The coincident and coherent multi-detector searches for gravitational waves, which by itself is a computing resources intensive work, is based on template banks of model signals computed for the variety of astrophysical sources and events producing gravitational wave signals. Such model signals are the result of computing intensive numerical relativity calculations for the various gravitational phenomena, which often can only be performed on High-Performance Computing (HPC) facilities. Generally speaking HPC computing facilities, which are optimized for processing speed, are important for the astrophysical and cosmological modelling.

For dark energy surveys where large numbers of astronomical images are taken, these images are being calibrated with sets of calibration images taken together with the actual observations. Depending on the character of such images (direct imaging of the sky or images with spectra taken by a spectrograph at a telescope) a further processing allows to extract individual sources or bin the image information in spectra. The reduction tasks are often automated by a data reduction pipeline allowing to process large numbers of images automatically.

Common for most of the astroparticle data analyses is that these are performed on High-Throughput Computing (HTC) systems, which are large clusters of computing resources optimized for data processing throughput. These HTC systems make use of multiprocessor hardware units; increasingly GPU based systems are installed. To make full use of the computing power provided by such systems for scientific data analysis a special middleware is required. The gravitational wave community has been first in using GPU bases systems for their data analysis. Other communities are now using GPU based systems as well, e.g. for detailed extensive air showers simulations.

Computing services used for astroparticle data analysis are based either on dedicated computing infrastructures or on the WLCG Grid infrastructure, which was originally set up for high energy physics data analysis. Another future possibility may become available by making use of Cloud based services.

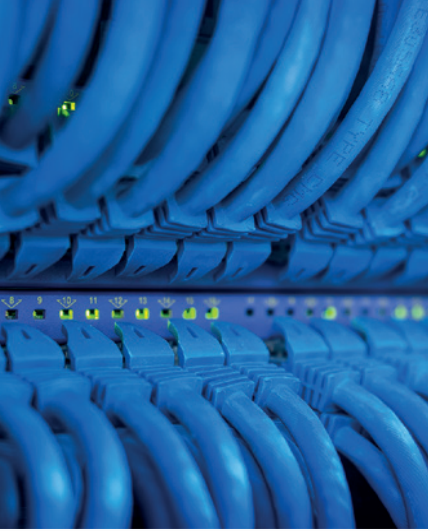
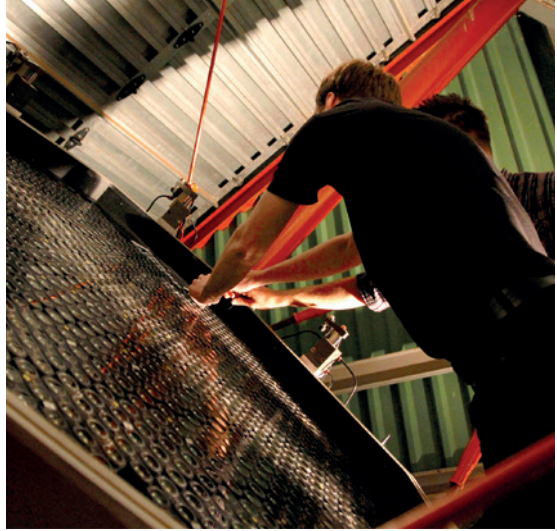
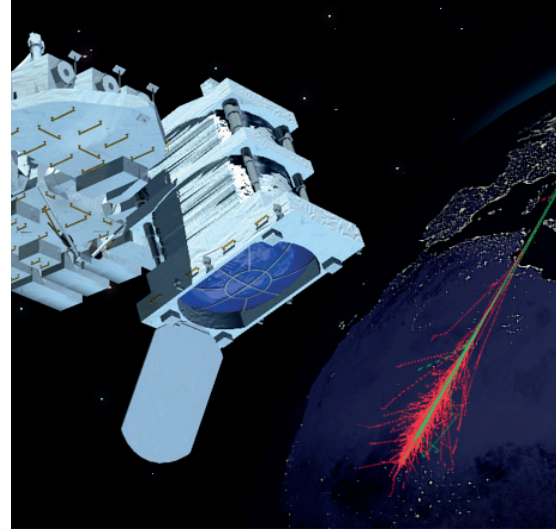

## 3.2 Data storage and distribution

**Data storage**

In most of the European research data centers the managing of data from astroparticle physics projects is one among other tasks. Among the data centers listed in Table 1 the Albert Einstein Institute (AEI) in Hanover is the only center highly specialized on managing astroparticle physics data. The last years have demonstrated that the combination of both solutions defines an efficient way of making use of existing resources and, when justified by a particular scientific computing demand, setup a dedicated data center.

As far as data storage is concerned, for high energy cosmic ray experiments the overall data storage requirement typically divides into a ratio of raw data : reconstructed data : simulation data of about 60:10:30. For ground-based gravitational wave experiments data storage is dominated by the raw data. Figure 1 indicates that the demand in both, short and long term storage of major current and future astroparticle physics projects is increasing by about a factor of 10 till the end of this decade. This is of course well-founded by the next generation of projects with substantially improved detector technology providing much higher resolution data. Furthermore, much more advanced analysis possibilities and methods will contribute to the overall increasing data production in the future. This is not only the case for astroparticle physics projects it defines a general trend in science. To cope with the increasing demand in data storage by astroparticle physics projects new data analysis methods should become integral part of the extension strategies and plans of all data centers involved.

**Data distribution**

The distribution of main astroparticle physics data over European data centers tabulated in Table 1 shows that in most cases data of an individual project are stored in more than one research data center. Note that in several cases (e.g. IceCube, LIGO, Fermi) further research data centers outside Europe are involved. This increases the level of data safety and would allow sharing the data access. On the other hand in several research data centers project data of the same type of projects are managed, which is efficient in a way that more than one project can benefit from local expertise and common analyses can be carried out without transferring larger amounts of data. Figure 3 demonstrated the complexity of the distribution of astroparticle physics data stored in Europe and the individual way how these data are transferred from the individual project sites to the centers. This is vastly different compared to the WLCG where data taken by the CERN LHC detectors are transferred from the central Tier-0 center to a number of Tier centers sharing the storage and processing of the entire data. However, in both cases the location of individual data sets is exactly defined, Compared to the WLCG only a common interface to get access to all the astroparticle physics data is not available yet.

**Database systems**

Relational databases like Oracle, MySQL, or PostgreSQL are used by most of the astroparticle physics projects. VIRGO, LIGO and LSST for example are storing data in MySQL databases whereas IceCube is using MySQL as repository for calibration data.

Since the scaling of the current relational database systems is limited, future projects may need to consider other database technologies. NoSQL (Not only SQL) databases, which are increasingly used in big data and real-time web applications, may provide a solution to overcome the limitations of purely relational databases. Developments for these new technologies are obviously pushed by the needs of the commercial big data sector. This should be followed by the scientific community. It would be desirable that research data centers and scientific community – when going in this direction – could agree on standards for database systems allowing handling large distributed data set on a distributed computing infrastructure.

## 3.3 Grid vs. Cloud technology

As already mentioned several astroparticle projects have used resources of the WLCG for parts of their simulations and analysis, taking advantage of the grid infrastructure set up for high energy physics data analysis. This Grid runs today about 1,000,000 jobs per day, a number demonstrating that data analysis in high energy physics is largely batch, either for large Monte Carlo simulations or for event reconstruction. Another example of a Grid infrastructure is the LSC Data Grid (LIGO Data Grid), which provides a coherent and uniform LIGO data analysis environment by combining computational and data storage resources of the LIGO Scientific Collaboration with grid computing middleware. In Europe the Albert Einstein Institutes in Hanover and Golm, and the Universities of Birmingham and Cardiff are providing computing resources to the LIGO Data Grid. The LIGO Data Grid is widely used for a combined data analysis of all current gravitational wave experiments.

Another concept of converged infrastructures and sharing services over a network is provided by the cloud computing (often just called the Cloud). It is worthwhile to mention that the WLCG already makes use of some Cloud techniques. Cloud computing allows for several basic services, namely:

- SaaS (Software as a Service) – the software is available on remote computers, and data can be either local or remote;
- PaaS (Platform as a Service) – complete software platforms are available on remote data-centers, with a specific platform e.g. for each astroparticle project;
- DaaS (Data as a Service) – the data are available on remote computers, and can be accessed either locally or remotely;
- HaaS (Hardware as a Service) – the user has both data and software, and send them to remote computers to run the jobs;
- IaaS (Infrastructure as a Service) – very similar to Grid Computing, but resources are used ondemand, without the need to pledge them, as is the case for several LHC experiments.

The underlying technologies have been mainly developed by large internet companies (e.g. Google and Amazon) that make use of the Cloud for their own purposes, to offer products and services to their customers. During the last years it became a business on its own for such companies to sell their Cloud based resources and services. Customer can choose between different business models such as a pay-per-use basis or paying a subscription fee. Buying Cloud services from a company might be an option to handle a temporary demand in computing resources to e.g. cache large amounts of data before they can handled by a scientific data center.

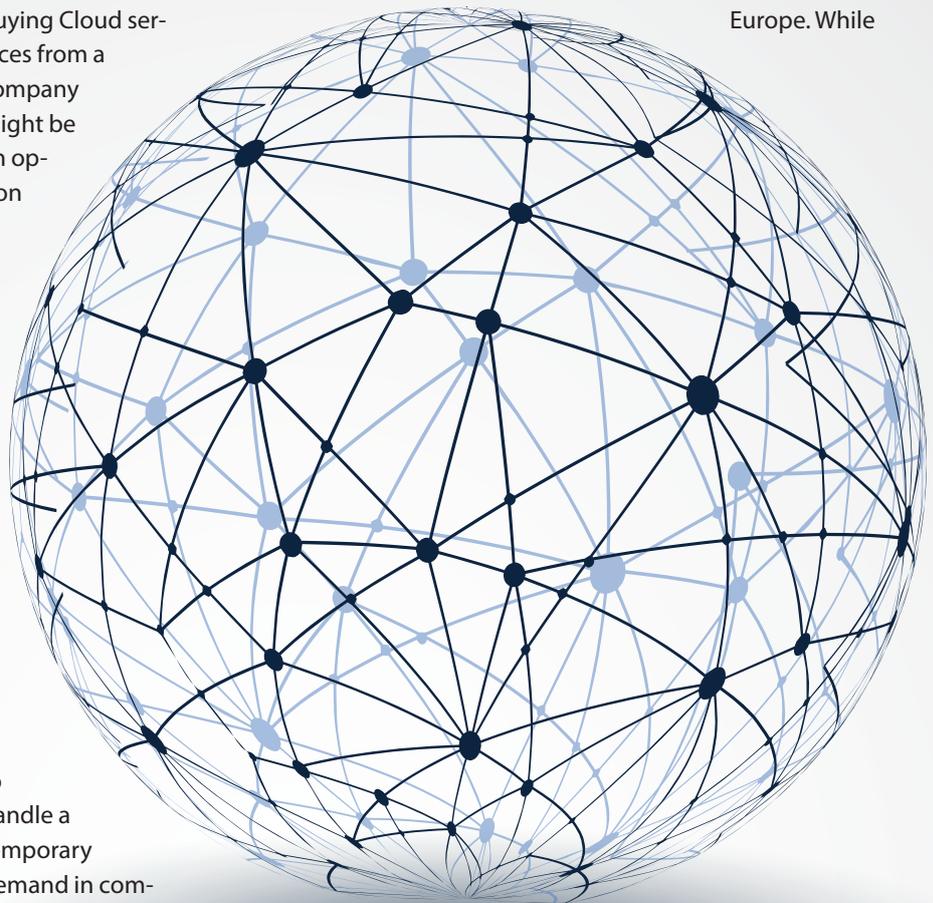

More important for scientific data analysis is however that Cloud technologies can also be used to setup private Cloud infrastructure in any data center; many scientific data centers, research institutes and universities have already installed private Clouds. Several initiatives exist to interconnect these individual Clouds to a federated Cloud system. For instance, OpenCloudMesh[6] is a joint international initiative of research institutes under the umbrella of the GÉANT Association to provide a universal file access beyond the borders of individual Clouds and into a globally interconnected mesh of research clouds. From the computing centers listed in Table 1, currently CERN and DESY are involved in this initiative. Another initiative that shall be mentioned is the Helix Nebula[7], which started as a European Commission-funded project to define the concept of a federated Cloud system for science in Europe. While partners of this project from academia (e.g. CERN, ESA, EMBL) and industry continue to collaborate in the Helix Nebula initiative, the implementation of a European Science Cloud is still an open issue.

It can be summarized that a future Cloud infrastructure between European scientific data centers would allow accommodating the needs of the different scientific community including astroparticle physics.

[6] https://owncloud.com/lp/opencloudmesh/

[7] http://www.helix-nebula.eu

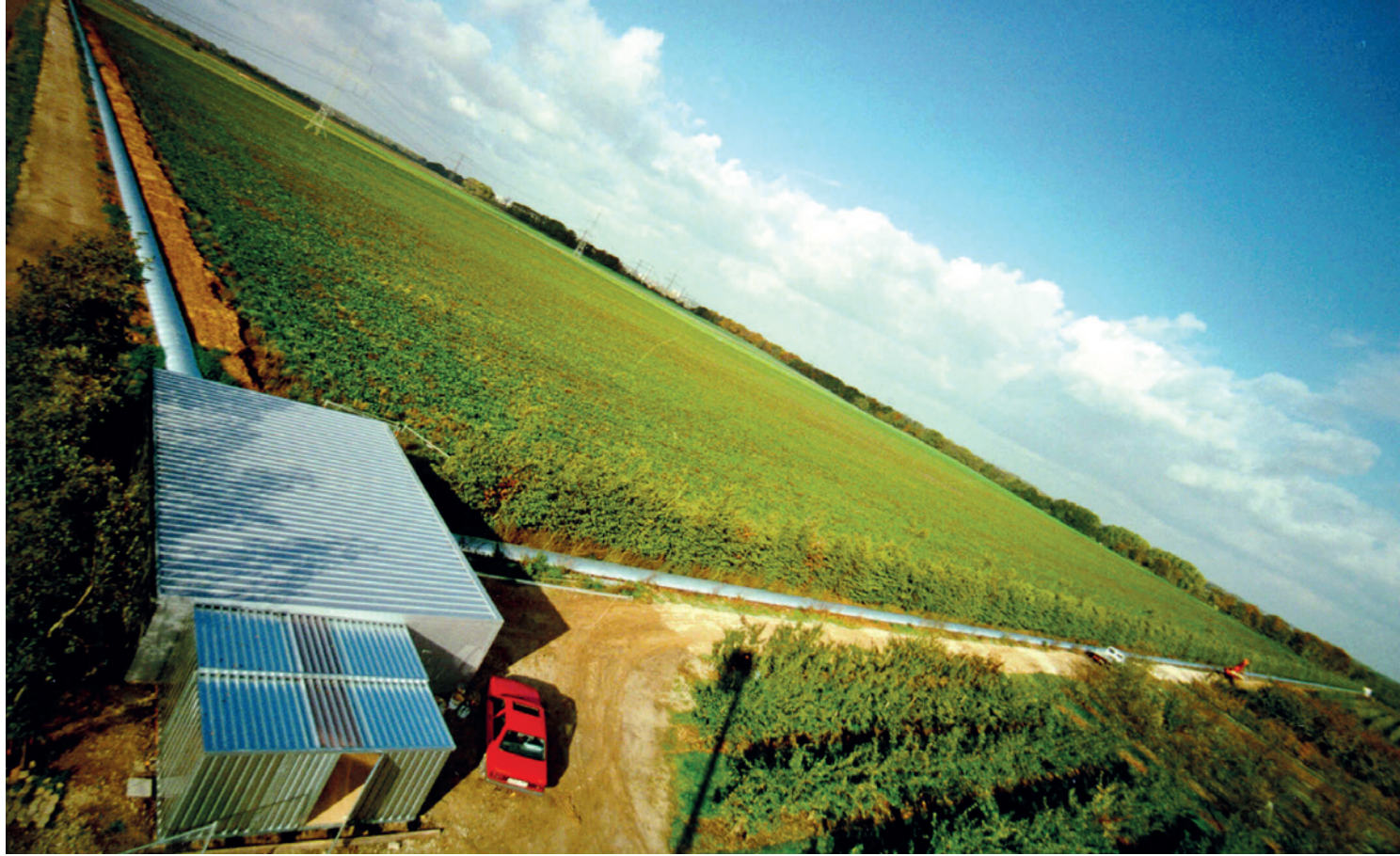

## 3.4 Data preservation

A major task in managing astroparticle physics data – as well as for any other important scientific data – is to securely store the data in a suitable number of repositories, and ensure its long term readability. Given the increasingly large amounts of data it is of equal importance to define an intelligent geographical distribution of data and its repositories in a way that data traffic can be handled in the future.

As far as a multi-messenger analysis with astroparticle physics data is concerned it seems quite natural to bring together the results of the individual astroparticle physics projects in the frame of the Virtual Observatory (VO). This may become a central preoccupation of the future large infrastructures in astroparticle physics. The Virtual Observatory is an international community-based initiative in astrophysics. It aims to allow global electronic access to the available astronomical data archives of space and ground-based observatories and other sky survey databases.

Cooperation between the stakeholders on the coordination level as well as the working level is necessary to achieve the integration of astroparticle physics data in the VO, which might be offering an open access to astroparticle physics data. Nevertheless, the project collaborations, large data centers, and funding agencies must develop solutions for managing the astroparticle physics data.

In view of long-term data preservation additional aspects are to be considered (see e.g. Diaconu et al. 2014[8], 2015). The data themselves are going to be useful only in case also documentation, software, and especially expertise and knowledge about the experiments are going to be preserved. In terms of preservation of software, the virtualization of the operating system and analysis software in virtual machines can enable us to maintain the ability to run specific software beyond the availability of computing hardware infrastructures. In view of this, a common archive of virtual machines and the dedicated software would be of large value for the community in its effort to provide the possibility of re-analyzing and re-investigating data of past experiments.

[8] Diaconu et al. 2014, "Scientific Data Preservation 2014", CNRS publication, http://hal.in2p3.fr/in2p3-00959072

# 4. Accessing astroparticle physics data, the researcher's view

Concerning the organization of an access to astroparticle physics data there are different aspects to consider. To enable researchers a scientific use of the data generated by astroparticle physics experiments technical solutions have to be provided to e.g. allow for data mining in larger data sets, multi-messenger analyses, modelling and simulations.

At present in running astroparticle physics projects there are a variety of user analysis environments in use, which have been developed in the context of the Grid or other high-throughput computing surroundings. Concerning data that can be mapped to the sky the Virtual Observatory would provide another environment for organizing data access. In the future the demands in user analysis environments will definitely grow with the number of users, the amount of generated data, and the number of data centers involved. The technical aspects are provided in the following Section 4.1. Section 4.2 shall provide a view on policy aspects related to the access to data from astroparticle physics projects.

## 4.1 User analysis environments

Currently, users of data from astroparticle physics experiments use several different analysis environments, depending on the type of data users are analyzing and depending on preferences of users as well as their knowledge and training.
In particle physics, where event-type data are most common, it has become obvious that for the end user analysis special computing farms with local storage for the users are necessary, e.g. the NAF at DESY or various WLCG Tier-3 centers. In these end user analysis centers batch submission systems are installed to allow the users to do their data analysis. Batch submission is handled with systems like Grid Engine or HTCondor. Root/Python is often used as user analysis environment. Currently there are of the order of 3000 users. This number is expected to at least double till 2020 and – under today's conditions – require extra (temporary) disk space of the order of 6 PBytes, which could be realized by different resource centers like Grid sites or cloud providers. The extra CPU usage is at present difficult to estimate.

Analysis environments exist at CC-IN2P3 and CNAF allowing carrying out an analysis of the Virgo data. The LIGO Data Grid provides a different analysis environment for users working with time-series obtained with LIGO and GEO600. Most of the Scientific data analyses however require a combined analyses of Virgo, LIGO and GEO600 data, which is to a large extend carried out in the frame of the LIGO Data Grid and its software tools.

For image-type data the astronomy community and observatories have developed a large variety of calibration and analysis software systems, which allows users to interactively work with such data or can be implemented in data reduction pipelines to produce standard products for larger sets of input data.

Another tool with growing popularity in many fields of science is the R software programming language and software environment for statistical computing and graphics, which has received a lot of commercial support in the context of big data analysis. R allows statisticians to do very sophisticated and complicated analyses with excellent graphics capabilities superior to other graphics/analysis packages commonly used e.g. in astronomy.

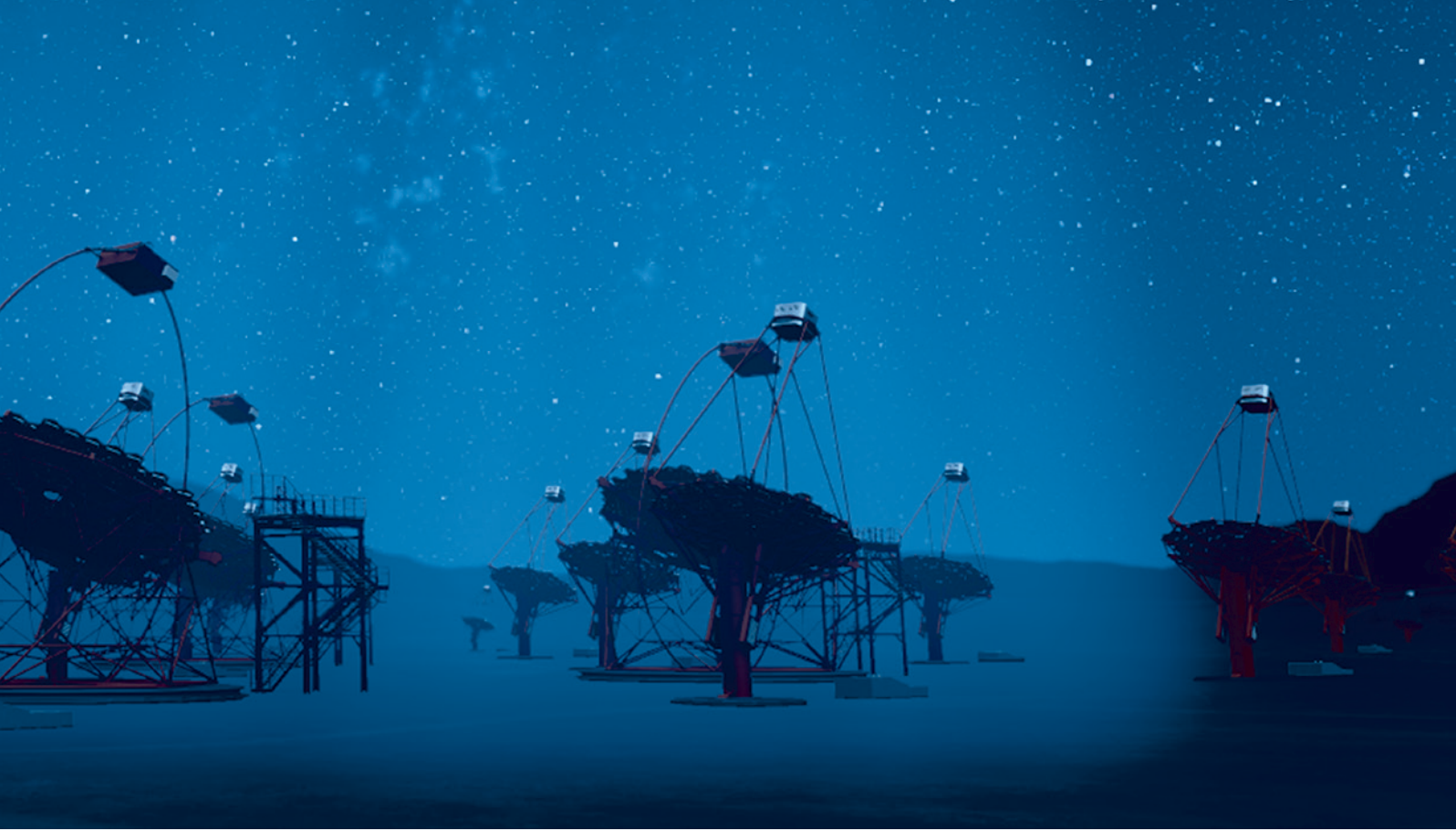

## 4.2 Data access policy

Research infrastructures like large astroparticle physics projects are built to provide the scientific community new possibilities in doing research. To balance between the interests of scientific groups largely participating in the construction, the interest of principal investigators (PIs) with their original research ideas, the interest of the entire scientific community for an archival access, and the interest of funding institutions investing in the construction and the running in best spending of public funding for research, research infrastructures develop data access policies. All major astroparticle physics projects have such access policies, which describe pathways to access project data. Accessing data of current astroparticle physics projects often requires being a member of the collaboration or at least an associate membership. In some cases a direct access to certain data products is possible via public data pages. For instance the PIC data center provides public MAGIC results through a Virtual Observatory server (http://vobs.magic.pic.es/) and releases public MAGIC results in FITS format.

With more and more scientists accessing astroparticle physics data the organization of an efficient open access to the scientific data needs to be carefully evaluated. For space born experiments (e.g. ESA satellites) and for astronomical observatories (e.g. ESO) it is common practice to grant PIs exclusive access to their scientific data. After a defined proprietary period– suitable for a scientific analysis by the PI – data are made publicly available via archives. When the technical challenges associated with an open access to astroparticle physics data are solved such existing policies for organizing data access can easily be adopted by coming astroparticle physics infrastructures.

While it is clear that each community can develop a solution for accessing their data, the ability of carrying out multi-messenger analyses with astroparticle physics data is becoming increasingly important. It seems quite natural to bring together the results of the individual astroparticle physics projects in a common frame such as the Virtual Observatory (VO). This may become a central preoccupation of the future large infrastructures in astroparticle physics. The Virtual Observatory is an international community-based initiative in astrophysics. It aims to allow global electronic access to the available astronomical data archives of space and ground-based observatories and other sky survey databases. Providing the framework for astroparticle physics products to be VO compliant means derive a:
- Data Model that provides a complete and shared descriptions of all data products.
- Data Access Layer that allows querying in a standard way all resources of all data providers.
- Data Exchange Format that allows one to retrieve data in a standardised way, including metadata.
- Information System Registry, describing the resources and services linked to astroparticle physics data, so making them available for any users. It is both a database and a web service holding metadata that describes all the resources.

Cooperation between the stakeholders on the coordination level as well as the working level is necessary to achieve the integration of astroparticle physics data in the VO, which then would offer an open access to astroparticle physics data. Nevertheless, the project collaborations, large data centers, and funding agencies shall develop together a solution for managing the astroparticle physics data.

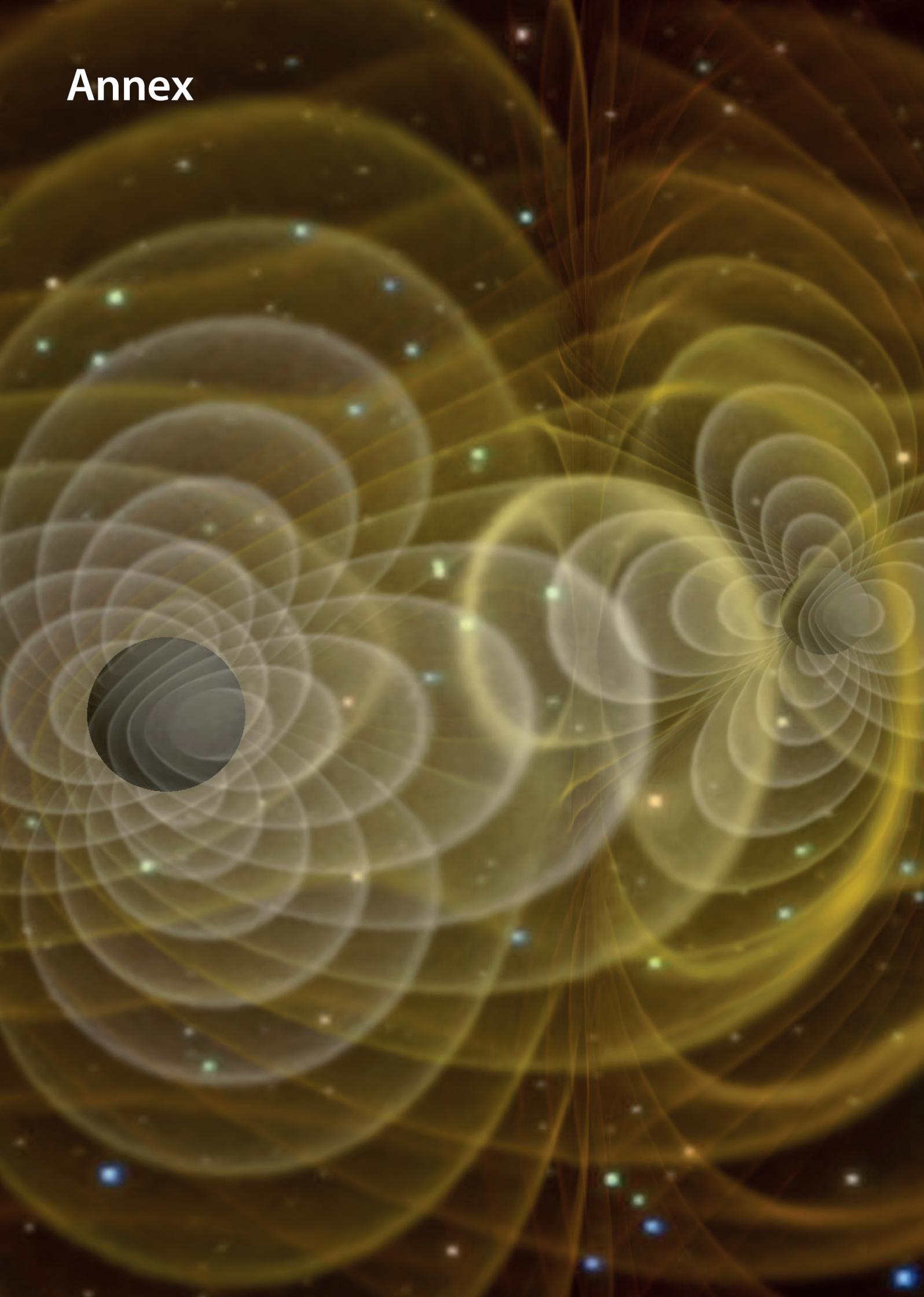

Annex

# A. CPU performance requirements of major current and future astroparticle physics projects

| | | 2011 | 2012 | 2013 | 2014 | 2015 | 2016 | 2017 | 2018 | 2019 | 2020 |
|---|---|---|---|---|---|---|---|---|---|---|---|
| data with event-like structure | AUGER (total) | 1,4 | 2,2 | 2,6 | 3,5 | 4,6 | 4,6 | 4,6 | 4,6 | 4,6 | 4,6 |
| | HESS | 1,6 | 2,0 | 10 | 14 | 16 | 16 | 16 | | | |
| | MAGIC | 0,8 | 0,8 | 8,0 | 8,0 | 8,0 | 8,0 | 8,0 | | | |
| | CTA | 10 | 10 | 10 | 10 | 10 | 10 | 24 | 30 | 40 | 50 |
| | ANTARES | 1,6 | 2,9 | 5,7 | 5,7 | 5,7 | | | | | |
| | KM3NeT | | | 0,5 | 0,5 | 3,0 | 10 | 20 | 50 | 100 | 200 |
| | IceCube | 15 | 15 | 15 | 15 | 15 | 15 | 15 | 15 | 15 | 15 |
| | FERMI | 15 | 15 | 15 | 15 | 15 | 15 | 15 | | | |
| | AMS | 2,3 | 9,0 | 30 | 50 | 60 | 70 | 80 | 90 | 100 | 110 |
| | *subtotal* | *48* | *57* | *97* | *122* | *137* | *149* | *183* | *190* | *260* | *380* |
| data with signal-like structure | VIRGO | 2,0 | 2,3 | 3,0 | 8,0 | 12 | | | | | |
| | advanced VIRGO | | | | | | 30 | 50 | 80 | 90 | 90 |
| | LIGO | | | 80 | 80 | 80 | | | | | |
| | advanced LIGO | | | | | | 750 | 1200 | 1200 | 1200 | 1200 |
| | *subtotal* | *2,0* | *2,3* | *83* | *88* | *92* | *780* | *1250* | *1280* | *1290* | *1290* |
| data with image-like structure | PAU | 10 | 40 | 110 | 90 | 90 | 90 | 90 | 90 | 90 | 90 |
| | SNLS | 1,0 | 1,0 | 1,0 | 1,0 | 1,0 | | | | | |
| | EUCLID | | | 0,2 | 0,5 | 6,2 | 13 | 19 | 25 | 31 | 38 |
| | LSST | | | 0,2 | 0,5 | 6,2 | 14 | 28 | 43 | 57 | 152 |
| | *subtotal* | *11* | *41* | *111* | *92* | *103* | *117* | *137* | *158* | *178* | *280* |
| | Other | 2 | 2 | 2 | 2 | 2 | 2 | 2 | 2 | 2 | 2 |
| | **CPU (kHS06) total** | **63** | **102** | **293** | **304** | **335** | **1047** | **1572** | **1629** | **1730** | **1951** |
| | **LHC Tier-0(2012)** | 0,096 | 0,16 | 0,45 | 0,47 | 0,51 | 1,6 | 2,4 | 2,5 | 2,7 | 3,0 |

All values are provided in units of kHS06, which is a commonly used unit in particle physics; 1 kHS06 equals to the performance of 100 CPU cores of today.

Data provided in FLOPS have been converted to kHS06 taking into account that 1Tflop (double-precision peak) equals to about 1 kHS06.

Given that in 2012 the LHC Tier-0 center at CERN had 650,000 cores installed the last row gives the total CPU performance required by astroparticle physics projects in units of the LHC Tier-0 center in 2012.

Whenever there are non-European partners only the European part is taken into account.

The Fermi requirements include only those coming from the Fermi Large Area Telescope (LAT) and do not include the Fermi Gamma-ray Burst Monitor (GBM) instrument.

# B. Low latency / disk storage needs of major current and future astroparticle physics projects

| | | 2011 | 2012 | 2013 | 2014 | 2015 | 2016 | 2017 | 2018 | 2019 | 2020 |
|---|---|---|---|---|---|---|---|---|---|---|---|
| data with event-like structure | AUGER (total) | 0,8 | 0,2 | 0,1 | 0,2 | 0,2 | 0,2 | 0,2 | 0,2 | 0,2 | 0,2 |
| | HESS | 0,0 | 0,2 | 0,2 | 0,3 | 0,3 | 0,3 | 0,3 | 0,3 | 0,3 | 0,3 |
| | MAGIC | 0,2 | 0,2 | 0,2 | 0,2 | 0,2 | 0,2 | 0,2 | 0,2 | 0,2 | 0,2 |
| | CTA | 0,3 | 0,3 | 0,5 | 0,9 | 1,0 | 1,0 | 4,0 | 10 | 21 | 28 |
| | ANTARES | 0,1 | 0,1 | 0,1 | 0,1 | 0,1 | 0,1 | 0,1 | | | |
| | KM3NeT | | | | | | 0,2 | 0,4 | 0,8 | 1,5 | 3,0 | 5,0 |
| | IceCube | 0,4 | 0,4 | 0,4 | 0,4 | 0,4 | 0,4 | 0,4 | 0,4 | 0,4 | 0,4 |
| | FERMI | 2,0 | 2,0 | 2,0 | 2,0 | 2,0 | 2,0 | 2,0 | | | |
| | AMS | 0,2 | 0,3 | 0,5 | 1,5 | 2,0 | 2,5 | 3,0 | 3,5 | 4,0 | 4,5 |
| | subtotal | 3,8 | 3,6 | 3,9 | 5,5 | 6,3 | 7,0 | 11 | 16 | 29 | 39 |
| data with signal-like structure | VIRGO | 0,7 | 0,7 | 0,7 | 0,7 | 0,8 | | | | | |
| | advanced VIRGO | | | | | | 1,7 | 1,8 | 2,8 | 3,2 | 3,8 |
| | LIGO | 1,4 | 1,4 | 1,4 | 1,4 | 1,6 | | | | | |
| | advanced LIGO | | | | | | 3,4 | 3,6 | 5,6 | 6,4 | 7,6 |
| | subtotal | 2,1 | 2,1 | 2,1 | 2,1 | 2,4 | 5,1 | 5,4 | 8,4 | 9,6 | 11 |
| data with image-like structure | PAU | 0,002 | 0,01 | 0,02 | 0,05 | 0,1 | 0,1 | 0,1 | 0,1 | 0,1 | 0,1 |
| | SNLS | 0,024 | 0,054 | 0,02 | 0,02 | 0,02 | | | | | |
| | EUCLID | | | 0,05 | 0,1 | 0,15 | 0,8 | 1,5 | 2,2 | 3,0 | 3,7 | 4,4 |
| | LSST | | | 0,05 | 0,1 | 0,15 | 0,5 | 0,8 | 1,2 | 1,6 | 6,5 | 16 |
| | subtotal | 0,026 | 0,164 | 0,24 | 0,37 | 1,4 | 2,4 | 3,5 | 4,7 | 10 | 21 |
| | Other | 1 | 1 | 1 | 1 | 1 | 1 | 1 | 1 | 1 | 1 |
| **Low Latency/Disk (PB) total** | | **7,0** | **6,9** | **7,3** | **9,0** | **11** | **16** | **21** | **30** | **50** | **72** |
| | LHC Tier-0 (2012) | 0,23 | 0,23 | 0,24 | 0,30 | 0,37 | 0,52 | 0,69 | 1,0 | 1,7 | 2,4 |

All values are provided in Petabyte (PB).

Given that in 2012 the LHC Tier-0 center at CERN provided the capability to store 30 PB on disk the last row gives the total low latency storage space required by astroparticle physics projects in units of the LHC Tier-0 center in 2012.

Whenever there are non-European partners only the European part is taken into account.

The Fermi requirements include only those coming from the Fermi Large Area Telescope (LAT) and do not include the Fermi Gamma-ray Burst Monitor (GBM) instrument.

# C. Archive / tape storage needs of major current and future astroparticle physics projects

| | | 2011 | 2012 | 2013 | 2014 | 2015 | 2016 | 2017 | 2018 | 2019 | 2020 |
|---|---|---|---|---|---|---|---|---|---|---|---|
| data with event-like structure | AUGER (total) | 0,2 | 0,4 | 0,4 | 0,4 | 0,5 | 0,5 | 0,5 | 0,5 | 0,5 | 0,5 |
| | HESS | 0,1 | 0,1 | 0,3 | 0,3 | 0,3 | 0,3 | 0,3 | 0,3 | 0,3 | 0,3 |
| | MAGIC | 0,8 | 0,8 | 1,2 | 1,7 | 2,1 | 2,1 | 2,1 | 2,1 | 2,1 | 2,1 |
| | CTA | 0,1 | 0,1 | 0,1 | 0,1 | 0,3 | 1,0 | 6,0 | 12 | 23 | 40 |
| | ANTARES | 0,2 | 0,2 | 0,3 | 0,3 | 0,3 | 0,3 | 0,3 | | | |
| | KM3NeT | | | | | 0,5 | 1,0 | 3,0 | 5,0 | 10 | 20 |
| | IceCube | 0,2 | 0,2 | 0,2 | 0,3 | 0,3 | 0,3 | 0,3 | 0,3 | 0,3 | 0,3 |
| | FERMI | 2,0 | 2,0 | 2,0 | 2,0 | 2,0 | 2,0 | 2,0 | 2,0 | 2,0 | 2,0 |
| | AMS | 0,0 | 0,0 | 0,5 | 2,0 | 2,0 | 2,0 | 2,0 | 2,0 | 2,0 | 2,0 |
| | subtotal | 3,5 | 3,7 | 4,9 | 7,0 | 8,2 | 9,4 | 16 | 24 | 40 | 67 |
| data with signal-like structure | VIRGO | 1,5 | 1,5 | 1,7 | 1,7 | 1,7 | | | | | |
| | advanced VIRGO | | | | | | 2,5 | 4,3 | 5,4 | 6,2 | 7,0 |
| | LIGO | 2,0 | 2,0 | 2,0 | 2,0 | 2,0 | | | | | |
| | advanced LIGO | | | | | | 6,0 | 6,8 | 7,6 | 8,4 | 9,2 |
| | subtotal | 3,5 | 3,5 | 3,7 | 3,7 | 3,7 | 8,5 | 11 | 13 | 15 | 16 |
| data with image-like structure | PAU | 0,01 | 0,03 | 0,10 | 0,10 | 0,30 | 1,3 | 2,3 | 3,3 | 4,3 | 5,3 |
| | SNLS | | | | | | | | | | |
| | EUCLID | | | 0,015 | 0,07 | 2,7 | 5,1 | 7,5 | 9,8 | 12 | 15 |
| | LSST | | | 0,05 | 0,05 | 0,05 | 0,60 | 1,1 | 1,6 | 2,6 | 5,6 | 19 |
| | subtotal | 0,01 | 0,08 | 0,17 | 0,22 | 3,6 | 7,5 | 11 | 16 | 22 | 39 |
| | Other | 1 | 1 | 1 | 1 | 1 | 1 | 1 | 1 | 1 | 1 |
| | **Tape storage (PB) total** | **8,0** | **8,2** | **9,8** | **12** | **17** | **26** | **40** | **54** | **78** | **123** |
| | **LHC Tier-0 (2012)** | 0,27 | 0,27 | 0,33 | 0,40 | 0,55 | 0,88 | 1,3 | 1,8 | 2,6 | 4,1 |

All values are provided in Petabyte (PB).

Given that in 2012 the LHC Tier-0 center at CERN provided the capability to store 30 PB on disk the last row gives the total archive (tape) storage space required by astroparticle physics projects in units of the LHC Tier-0 center in 2012.

Whenever there are non-European partners only the European part is taken into account.

The Fermi requirements include only those coming from the Fermi Large Area Telescope (LAT) and do not include the Fermi Gamma-ray Burst Monitor (GBM) instrument.

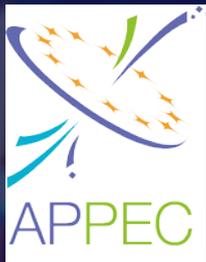

www.appec.org